\documentclass[aps,pra,nofootinbib,preprint,amsmath,amssymb,floatfix]{revtex4}
\usepackage{color}
\usepackage{graphicx}

\begin{document}

\newcommand{\tc}{\textcolor}
\newcommand{\g}{blue}
\newcommand{\ve}{\varepsilon}
\title{Casimir friction between a magnetic and a dielectric material, in the nonretarded limit}         

\author{Johan S. H\o ye$^1$ and  Iver Brevik$^2$  }      
\affiliation{$^1$Department of Physics, Norwegian University of Science and Technology, N-7491 Trondheim, Norway}

\affiliation{$^2$Department of Energy and Process Engineering, Norwegian University of Science and Technology, N-7491 Trondheim, Norway}

\begin{abstract}

The repulsive nature of the static Casimir force between two half-spaces, one perfectly dielectric and the other purely magnetic, has been known since Boyer's work [T. H. Boyer, Phys. Rev. A {\bf 9}, 2078 (1974)]. We here analyze the corresponding friction force in the magnetodielectric case. Our main method is that of quantum mechanical statistical mechanics.  {The basic model we introduce} is a harmonic oscillator model: an electric dipole oscillating in the $x$ direction and a magnetic one oscillating in the $y$ direction, while their separation is in the $z$ direction. {This is then extended to particles with isotropic polarizabilities.} We evaluate the friction force in a variety of cases: force between moving particles, between a moving particle and a half-plane, and between half-spaces sliding against each other. {At the end explicit results are obtained both for finite and zero temperatures. We restrict ourselves to the nonretarded limit.}
\end{abstract}
\maketitle

\bigskip
\section{Introduction}
\label{secintro}

Casimir friction can take place between moving atoms, between an atom moving parallel to a plane surface at rest, or between closely spaced surfaces moving relative to each other. Being basically a non-equilibrium effect, it is nevertheless described usually in terms of the fluctuation-dissipation theorem, meaning that the two-point function for the field components are taken to be proportional to the imaginary part of the retarded Green function. This means physically that the electromagnetic field is assumed to be in local thermal equilibrium. Usually, magnetic properties of the material are left out, and the effect is accordingly due to fluctuating electric dipoles. Some  articles on the subject can be found in Refs.~\cite{milton16,silveirinha14,dedkov17,dedkov18,pendry97,barton10,barton11,intravaia14,intravaia15,intravaia16}; cf. also the extensive 2007 article \cite{volokitin07}.

If we move on to the case of magnetodielectric media, the situation becomes more involved. As shown first by Boyer \cite{boyer74} the static Casimir force between two parallel (thick) plates, the one being perfectly dielectric and the other perfectly magnetic, will be repulsive. We have recently contributed to the discussion on this kind of repulsive static forces, both with the  use  of macroscopic electrodynamics \cite{brevik18}, and from a statistical mechanical standpoint \cite{hoye18}.

The purpose of this paper is to analyze magnetodielectric {\it friction} between a particle and a half-space, as well as  between two half-spaces. This is a topic not discussed earlier in the literature as far as we know. As one would expect, the friction force will turn out to be very small, actually much smaller than in the standard nonmagnetic case. The force must therefore be regarded as an esoteric quantity, far beyond measurability by present experimental techniques. The significance of the force lies in its existence, not in its magnitude. An important point worth noticing is, however, that it always acts so as to oppose the motion of the plates, just as friction always behaves in ordinary hydrodynamics.

Our basic method will be the one  of quantum statistical mechanics. In the next section we introduce a simplified model for interacting magnetic and dielectric particles and consider then, in Sec.~III, a  harmonic oscillator model of an electric dipole and a magnetic dipole oscillating respectively in the $x$  and $y$ directions only while their separation is in the $z$ direction. The Hamiltonian is derived and the eigenfrequencies determined. In Sec.~IV, we look at the same oscillator model from another angle, namely by using the quantum mechanical path integral method. We find the results obtained from the two methods to agree. In Sec.~V we derive the friction force, via the Kubo formula (equivalent to the fluctuation-dissipation theorem) from which the response function needed is obtained. This  force turns out to have the proper sign for a braking force, but the force is much smaller than in the purely dielectric case. In Sec.~VI we extend our basic results to a pair of polarizable magnetic and dielectric particles, and consider in Sec.~VII  the friction between a particle and a half-space, and between two half spaces.

In Secs.~\ref{sec5}-\ref{sec7}, where finite temperatures are considered, the two kinds of particles are assumed to have sharp eigenfrequencies. This has as consequence that the friction force becomes a $\delta$-function in the frequency difference. This singular feature is smoothed out when the eigenfrequencies are replaced by continuous eigenfrequency spectra. This is included in Sec.~\ref{sec8}. And in Sec.~\ref{sec9} explicit results are obtained for a pair of half-spaces where both the dielectric as well as the magnetic frequency spectra follow the dielectric one of a metal. Results are then also derived and obtained for the zero temperature case. The forces obtained are vanishingly small, but can be related in a simple way to established explicit results where both half-spaces are dielectric.

Note: Much of the material reported in the following sections builds upon and  extends on previous papers of ours in this area throughout many years \cite{hoye18,hoye92,hoye10,hoye12,hoye13,hoye14,hoye93,brevik88}. Naturally, we have not found it possible to cover the various assumptions and derivations for each subtopic in detail   here, but we have attempted to give sufficient references for readers wishing to go into further detail.

{ We should also mention that we are ignoring retardation effects throughout. (Some investigators might therefore prefer to associate this kind of theory with van der Waals, instead of with Casimir.)}

We use Gaussian units throughout.

\section{Model for an interacting magnetic and a dielectric particle}
\label{sec2}

Magnetic and dielectric particles interact via the radiating electromagnetic field. When a pair of these particles interact at thermal equilibrium it has been found, as mentioned above, that  the induced force is repulsive, thus  in contrast to the usual Casimir force \cite{hoye18,boyer74}. This is not immediately obvious on physical grounds.
 However some physical insight can be obtained, as shown in Ref.~\cite{hoye18}, by constructing a model with 3 oscillators: two of them interacting with the third one, the latter playing the role of the electromagnetic field. We show now that this construction can be simplified a bit further by use of the quasistatic fields created by electric currents.

Specifically, the quasistatic model may consist of an electric dipole where a charge oscillates in a given direction. The magnetic dipole may be a current oscillating in a circular loop. The current in the electric dipole will create a magnetic field that interacts with the magnetic dipole moment. Conversely the changing magnetic field, produced by the current loop by electromagnetic induction, induces an electric field that interacts with the electric dipole.

By solution of Maxwell's equations the magnetic field ${\bf H}$ induced by an oscillating electric dipole moment ${\bf P}$ is given by Eq.~(28) of Ref.~\cite{hoye18} as
\begin{equation}
{\bf H}=-\zeta(1+\zeta r)\frac{e^{-\zeta r}}{r^2}(\hat{\bf r}\times{\bf P}).
\label{10}
\end{equation}
 Here Fourier transform with respect to time has been performed such that ${\bf H}$ and ${\bf P}$ mean Fourier transformed quantities, and
\begin{equation}
\zeta=i\frac{\omega}{c}
\label{11}
\end{equation}
where $\omega$ is frequency and $c$ is velocity of light. The ${\bf r}$ is spatial separation, and the hat denotes unit vector. For small   $\zeta r$ in the quasistatic limit expression (\ref{10}) simplifies to
\begin{equation}
{\bf H}=-\zeta\frac{\hat{\bf r}\times{\bf P}}{r^2}.
\label{12}
\end{equation}
Transforming back to time dependence this becomes ($i\omega\rightarrow\partial/\partial t$, $t$ is time)
{
\begin{equation}
{\bf H}=\frac{\dot{\bf P}\times\hat{\bf r}}{cr^2}=\frac{d{\bf l}\times\hat{\bf r}}{cr^2}I.
\label{13}
\end{equation}
}
where $I$ is the electric current in a wire element of length and direction $d{\bf l}$. Expression (\ref{13}) is the Biot-Savart's law for the magnetic field created by a current element where $\hat{\bf r}$ is the unit vector for the direction in space. The direction of  $\hat{\bf r}$ is from the current element toward the point of observation.

Due to symmetry of Maxwell's equations the electric field ${\bf E}$ created by an oscillating magnetic dipole ${\bf M}$ also follows from Eq.~(\ref{10}) by removing the minus sign in front. However, the unit vector $\hat{\bf r}$ may still be directed from the electric to the magnetic dipole. This will restore the minus sign, and for small $\zeta r$ the equation corresponding to Eq.~(\ref{13}) becomes
\begin{equation}
{\bf E}=\zeta\frac{\dot{\bf M}\times\hat{\bf r}}{cr^2}
\label{14}
\end{equation}

With Eqs.~(\ref{13}) and (\ref{14}) we now apparently obtain two different expressions for the energy of interaction. These are
\begin{equation}
-\Delta L_H=- {\bf HM}=-2\alpha(\dot{\bf P}\times\hat{\bf r}){\bf M} \quad \mbox{and}\quad -\Delta L_E=- {\bf EP}=-2\alpha(\dot{\bf M}\times\hat{\bf r}){\bf P}.
\label{15}
\end{equation}
with $\alpha=1/(2cr^2)$. {  (Later, in Sec. VI, we will let the same symbol $\alpha$ stand for a polarizability.) }
As we will find, these expressions are consistent when used to obtain the equations of motion.

\section{Equations of motion}
\label{sec3}

Consider the simple harmonic oscillator model of an electric dipole oscillating in the $x$ direction and a magnetic one oscillating in the $y$ direction while their separation is along the $z$ direction. With new coordinates $P\rightarrow x$ and $M\rightarrow y$ with $\hat{\bf r}$ pointing in the positive $z$ direction, interaction (\ref{15}) becomes
\begin{equation}
\Delta L_H=-2\alpha \dot{x}y\quad \mbox{and}\quad \Delta L_E=2\alpha x\dot{y}.
\label{16}
\end{equation}

Assume for simplicity that the non-interacting system is two harmonic oscillators with same eigenfrequency $\omega_0=1$. Their Lagrange function is then
\begin{equation}
L_0=\frac{1}{2}(\dot{x}^2+\dot{y}^2)-\frac{1}{2}(x^2+y^2).
\label{17}
\end{equation}
Contributions (\ref{16}) turn out to be equivalent, and we take half of each to obtain the resulting Lagrange function
\begin{equation}
L=L_0+\frac{1}{2}(\Delta L_H+\Delta L_E)
\label{18}
\end{equation}
[Other combinations by adding a term $\rm{const.}(\Delta L_H-\Delta L_E)$ {do} not change the dynamics since $\dot{x}y+x\dot{y}=d(xy)/dt$.]

From Eqs.~(\ref{16})-(\ref{18}) one finds the generalized momenta
\begin{equation}
p_x=\frac{\partial L}{\partial\dot{x}}=\dot{x}-\alpha y\quad\mbox{and}\quad p_y=\frac{\partial L}{\partial\dot{y}}=\dot{y}+\alpha x.
\label{19}
\end{equation}
With the variational principle the classical equations of motion follow from Lagrange's equations
\begin{eqnarray}
\nonumber
\dot p_x-\frac{\partial L}{\partial x}=\ddot{x}+x-2\alpha\dot{y}=0\\
\dot p_y-\frac{\partial L}{\partial y}=\ddot{y}+y+2\alpha\dot{x}=0.
\label{20}
\end{eqnarray}
With solution of form $x\sim  y \sim e^{i\omega t}$ the equation for the two eigenfrequencies becomes
\begin{equation}
(-\omega^2+1)^2-(-2\alpha i\omega)(2\alpha i\omega)=0.
\label{21}
\end{equation}
From this follows $-\omega^2+1=\mp 2\alpha\omega$ by which the eigenfrequencies ($>0$) are
\begin{equation}
\omega_\pm=\pm\alpha+\sqrt{1+\alpha^2}.
\label{22}
\end{equation}

With this result the ground state energy of the quantized harmonic oscillator system is
\begin{equation}
E_0=\frac{1}{2}\hbar(\omega_++\omega_-)=\hbar\sqrt{1+\alpha^2}.
\label{23}
\end{equation}
Accordingly at temperature $T=0$ there is a repulsive Casimir force between the two oscillators (provided $\alpha$ decreases by increasing separation).

At large temperatures the classical limit is obtained with internal energy $k_B T$ for each oscillator where $k_B$ is Boltzmann's constant. This is independent of couplings and eigenfrequencies of the oscillators. The entropy and free energy of the system have corresponding temperature dependent contributions. {In addition} the entropy and thus the free energy have a contribution that depends upon the logarithm of the product of eigenfrequencies. In the present case with eigenfrequencies (\ref{22}) one finds that this product $\omega_+\omega_-=1$ does not depend upon the interaction. Thus in the classical high temperature limit the resulting free energy does not depend upon $\alpha$. Therefore the induced interaction vanishes too, consistent with  the result of Ref.~\cite{hoye18}.

To obtain the Schr\"{o}dinger equation for the two coupled oscillators, one needs the Hamiltonian
\begin{equation}
H=p_x\dot{x}+p_y\dot{y}-L.
\label{24}
\end{equation}
With Eqs.~(\ref{16})-(\ref{18}) one finds
\begin{equation}
H=\frac{1}{2}\left[(p_x+\alpha y)^2+(p_y-\alpha x)^2+x^2+y^2\right].
\label{25}
\end{equation}
Solution of the Schr\"{o}dinger equation recovers the spectra of two harmonic oscillators with the eigenfrequencies (\ref{22}).

\section{Induced interaction}
\label{sec4}

Due to the special form of the Lagrange function for the model studied, it is of interest to verify that its evaluation via the path integral is consistent with the result of Sec.~\ref{sec3}. With imaginary time
\begin{equation}
\lambda=i\frac{t}{\hbar}, \quad \frac{1}{dt}=\frac{i}{\hbar}\frac{1}{d\lambda},
\label{30}
\end{equation}
the Lagrangian (\ref{16})-(\ref{18}) becomes
\begin{equation}
-L=\frac{1}{2}\left[\frac{1}{\hbar^2}\left(\left(\frac{dx}{d\lambda}\right)^2
+\left(\frac{dy}{d\lambda}\right)^2\right)+x^2+y^2\right]+\alpha\frac{i}{\hbar}\left[\frac{dx}{d\lambda}y-x\frac{d y}{d\lambda}\right].
\label{31}
\end{equation}
With Fourier transformed quantities
\begin{equation}
\tilde x(K)=\frac{1}{\sqrt{\beta}}\int\limits_0^\beta x(\lambda)e^{iK\lambda}\,d\lambda, \quad x(K)=\frac{1}{\sqrt{\beta}}\sum\limits_{n=-\infty}^{\infty} \tilde x(K)e^{-iK\lambda}
\label{32}
\end{equation}
and likewise for $\tilde y(K)$ one finds
\begin{equation}
\int\limits_0^\beta L\, d\lambda=\sum\limits_{n=-\infty}^{\infty}\left\{-\frac{1}{2}\left[\left(\frac{K}{\hbar}\right)^2+1\right]\left[\tilde x(K)\tilde x(-K)+\tilde y(K)\tilde y(-K)\right]+\Delta L_K\right\},
\label{34}
\end{equation}
\begin{equation}
\Delta L_K=-\alpha\frac{i}{\hbar}\left[-iK\tilde x(K)\tilde y(-K)-\tilde x(K)(iK\tilde y(-K))\right]=-2\alpha\frac{K}{\hbar}\tilde x(K)\tilde y(-K).
\label{35}
\end{equation}
The $K$ ($=i\hbar\omega$) are the Matsubara frequencies
\begin{equation}
K=\frac{2\pi n}{\beta}
\label{36}
\end{equation}
with $n$ integer and $\beta=1/(k_B T)$ where $k_B$ is Boltzmann's constant.

Now $\tilde x(-K)=\tilde x(K)^*$ by which one can introduce real variables such that {$\tilde x(K)=(b_x(K)+ic_x(K))/\sqrt{2}$ etc. by which} $K$ can be restricted to $K>0$ by adding together terms $\pm K$. We get
\begin{equation}
\Delta L_K+\Delta L_{-K}=-2\alpha i\frac{K}{\hbar}[c_x(K) b_y(K)-b_x(K) c_y(K)].
\label{37}
\end{equation}
In the path integral the exponential of expression (\ref{34}) is integrated with respect to $b_x(K)$, $c_x(K)$ etc. that form Gaussian integrals.

With neglect of $\Delta L_K$ one is left with the reference system of two independent oscillators where for the present case {one} finds the averages
\begin{equation}
\langle|\tilde x(K)|^2\rangle=\langle|\tilde y(K)|^2\rangle=\frac{1}{u^2+1}, \quad u=\frac{K}{\hbar}.
\label{38}
\end{equation}

For a given value of $K$ the perturbation gives a contribution $F_K$ to the free energy. With {(\ref{35}) and} (\ref{38}) this is
{
\begin{equation}
-\beta F_K=\frac{1}{2}\langle \Delta L_K \Delta L_{-K}\rangle=-2\alpha^2\frac{u^2}{(u^2+1)^2}.
\label{39}
\end{equation}
}
 The total contribution follows by summation over $K=2\pi n/\beta$. At $T=0$ ($\beta\rightarrow\infty$) one can integrate, and with $du=dK/\hbar=2\pi/(\hbar\beta)\,dn$ one obtains the free energy contribution
{
\begin{equation}
F=\sum\limits_{n=-\infty}^\infty F_K \rightarrow \frac{2\alpha^2 \hbar}{2\pi}\int\limits_{-\infty}^\infty\frac{u^2\,du}{(u^2+1)^2}=\frac{1}{2}\hbar \alpha^2.
\label{40}
\end{equation}
Thus in terms of $\alpha$ the perturbation in result (\ref{23}) is recovered.} In the classical limit $\beta\rightarrow 0$ only the $K=0$ term contributes, but since $F_0=0$ the $F$ vanishes in this limit as concluded below Eq.~(\ref{23}).

{The study of the interaction between dielectric and magnetic media initiated by Boyer \cite{boyer74} has given rise to later studies on the magnetodielectric interaction. In addition to the papers \cite{dedkov17,brevik18,hoye18} referred to  above, we mention that of Zhao {\it et al.} dealing with repulsive static Casimir forces in chiral metamaterials \cite{zhao09}, and  that of Nesterenko and Nesterenko dealing with Casimir friction \cite{nesterenko14}.}

\section{Casimir friction}
\label{sec5}

Earlier we have obtained results for the friction when interactions are instantaneous like the electrostatic interaction \cite{hoye92,hoye10,hoye12,hoye13,hoye14,milton16}. Also some results have been obtained for interactions that vary with time, but general results are less worked out and the situation is less clear \cite{hoye93}. With the present model where electric and magnetic dipole moments interact the situation is intermediate {with quasistatic interactions.} This simplifies, and as we will find below, our results for static interactions can be generalized. Details of explanations and derivations can thus be found in our earlier works.

Earlier we studied the friction for the basic system of two harmonic oscillators with coordinates $x_1$ and $x_2$. They interacted via the energy
\begin{equation}
-Aq(t)=\psi({\bf r}(t))x_1 x_2
\label{50}
\end{equation}
as given by {Eq.~(I1)} of Ref.~\cite{hoye10}. Here and below the numeral I is used to designate the equations of this reference. The $\psi({\bf r}(t))$ is the coupling strength which varies due to the relative motion with velocity ${\bf v}$. The $A$ is an operator since $x_1$ and $x_2$ are so in the quantum case while $q(t)$ is a classical function that gives the time dependence.

In the present case, however, the Hamiltonian is given by Eq.~(\ref{25}), and for small $\alpha$ the perturbing interaction is
\begin{equation}
-Aq(t)=\psi({\bf r}(t)) S, \quad S=\frac{1}{2}\left(\frac{p_x}{m_x}y-x\frac{p_y}{m_y}\right), \quad \psi({\bf r}(t))=2\alpha
\label{51}
\end{equation}
when the oscillators have masses $m_x$ and $m_y$ and also when they are extended to  have eigenfrequencies $\omega_x$ and $\omega_y$. (For small $\alpha$ this expression is mainly the perturbation of the Lagrangian too, since $p_x/m_x\approx \dot{x}$.)
Comparing one sees that the $x_1x_2$ part of interaction (\ref{50}) is replaced with the $S$ given by (\ref{51}). However, further evaluations can still follow closely those of Ref.~\cite{hoye10}. Thus Eqs.~(I2)-(I6) will be the same. The response function (I5) is needed to find the friction force. It is given by the Kubo formula \cite{kubo58,landau85,brevik88}
\begin{equation}
\phi_{BA}(t)=\frac{1}{i\hbar}\rm{Tr}\{\rho[A,{\bf B}(t)]\}
\label{52}
\end{equation}
where here $\rho$ is the density matrix. Only the time dependent part of the interaction $-A q(t)=\nabla \psi({\bf r}_0){\bf v}t$ {($\Delta{\bf r}={\bf r}-{\bf r}_0={\bf v}t$)} is needed. The ${\bf B}=\nabla\psi S$ is the force. Further with Eqs.~(I6) and (I7)
\begin{equation}
\phi_{BA}(t)={\bf G}\phi(t),\quad {\bf G}=(\nabla\psi)({\bf v}\cdot\nabla\psi), \quad \phi=\rm{Tr}\{\rho C(t)\}.
\label{53}
\end{equation}
But Eq.~(I7) is modified into
\begin{equation}
C(t)=\frac{1}{i\hbar}[S,S(t)]
\label{54}
\end{equation}
where $S(t)=e^{iHt/\hbar} S e^{-iHt/\hbar}$ is the Heisenberg operator with $H$ the unperturbed Hamiltonian. The expression for the friction force ${\bf F}_f$ is given by Eq.~(I9)
\begin{equation}
{\bf F}_f=-{\bf G}\int\limits_0^\infty\phi(u) u\,du,
\label{55}
\end{equation}
which we will use here too. [{Its} Fourier transformed version is Eq.~(I11).]

To evaluate the commutator $C(t)$ the harmonic oscillator annihilation and creation operators $a_i$ and $a_i^+$ are used for which
\begin{equation}
[a_i,a_i^+]=1, \quad a_i(t)=a_i e^{-i\omega_it}, \quad \quad a_i^+(t)=a_i^+ e^{i\omega_it}
\label{57}
\end{equation}
with $i=1,2$ since below we will replace coordinates $x$ and $y$ with  $x_1$ and $x_2$. Besides Eq.~(I12) for the operator $x_i$ we here will need the momentum operator $p_i$
\begin{equation}
x_i=\sqrt{\frac{\hbar}{2m_i \omega_i}}(a_i+a_i^+), \quad p_i=\frac{1}{i}\sqrt{\frac{\hbar m_i\omega_i}{2}}(a_i-a_i^+).
\label{58}
\end{equation}
In addition to Eq.~(I13)
\begin{equation}
L_i^+=L_i=\langle n_i|a_i a_i^+(t)+a_i^+ a_i(t)|n_i\rangle=(2n_i+1)\cos{(\omega_i t)}+i\sin{(\omega_i t)}
\label{60}
\end{equation}
we need
\begin{equation}
L_i^-=\langle n_i|a_i a_i^+(t)-a_i^+ a_i(t)|n_i\rangle=\cos{(\omega_i t)}+i(2n_i+1)\sin{(\omega_i t)}
\label{61}
\end{equation}

For the thermal average of $\phi(t)$ the result will be similar to (I14)
\begin{equation}
\phi(t)=\langle\langle n_1n_2|C(t)|n_1n_2\rangle\rangle=\frac{1}{i\hbar}\left(\frac{\hbar}{2}D\right) M, \quad D=\frac{\hbar}{2m_1m_2\omega_1\omega_2}.
\label{63}
\end{equation}

Based  on interaction (\ref{50}) Eq.~(I14) is obtained with $M=L_1 L_2-L_1^* L_2^*$. With the present interaction (\ref{51}) there will be 4 similar terms
\begin{eqnarray}
\nonumber
M&=&\frac{1}{4}[(\omega_1^2+\omega_2^2)(L_1^+ L_2^+-L_1^{+*} L_2^{+*})-2\omega_1\omega_2(L_1^- L_2^--L_1^{-*} L_2^{-*})]\\
&=&\frac{i}{2}\{(\omega_1^2+\omega_2^2)[(2\langle n_1\rangle+1)\cos{(\omega_1 t)}\sin{(\omega_2t)}+(2\langle n_2\rangle+1)\cos{(\omega_2 t)}\sin{(\omega_1t)}]
\label{64}\\
\nonumber
&&-2\omega_1\omega_2[(2\langle n_1\rangle+1)\cos{(\omega_2 t)}\sin{(\omega_1t)}+(2\langle n_2\rangle+1)\cos{(\omega_1 t)}\sin{(\omega_2t)}]\}.
\end{eqnarray}
Integral (I16) was needed to obtain the friction force
\begin{equation}
\int\limits_0^\infty te^{-\eta t}\cos{(\omega_1t)} \sin{(\omega_2t)}\, dt\rightarrow-\frac{\pi}{2\Omega}\delta(\Omega), \quad \eta\rightarrow 0
\label{65}
\end{equation}
where $\Omega=\omega_1-\omega_2$. The factor $e^{-i\eta t}$ is needed for convergence. Since only $\Omega\rightarrow 0$ contributes for the case of small $v$ at finite temperature a factor $(\omega_1-\omega_2)^2=\Omega^2$ can be neglected in (\ref{64}) by which it simplifies to
\begin{equation}
M=i\omega_1 \omega_2[(2\langle n_2\rangle+1)-(2\langle n_1\rangle+1)]\sin{(\Omega t)}.
\label{66}
\end{equation}
With this Eq.~(\ref{65}) is replaced by
\begin{equation}
\int\limits_0^\infty te^{-\eta t}\sin{(\Omega t)}\,dt\rightarrow \frac{\pi}{\Omega}\delta(\Omega), \quad \eta\rightarrow 0.
\label{67}
\end{equation}

At thermal equilibrium $2\langle n_i\rangle+1=\coth{(\beta\hbar\omega_i/2)}$ by which Eq.~(I18) will be as before
\begin{equation}
\coth{(\frac{1}{2}\beta\hbar\omega_1)}-\coth{(\frac{1}{2}\beta\hbar\omega_2)}\rightarrow\frac{\frac{1}{2}\beta\hbar\Omega}{\sinh{^2(\frac{1}{2}\beta\hbar\omega_1)}}, \quad \omega_2\rightarrow \omega_1.
\label{68}
\end{equation}
{Finally, for finite temperature} the friction force (\ref{55}) is obtained by multiplying Eq.~(\ref{68}) with (\ref{67}), the factor $-i\omega_1\omega_2$ from (\ref{66}), the $1/(2i)$ and $D$ from (\ref{63}), and the ${\bf G}$ from (\ref{53}). By that result (I19) multiplied with $\omega_1\omega_2\rightarrow \omega_1^2$ is obtained
\begin{equation}
{\bf F_f}=-\frac{\pi\beta\hbar^2(\nabla\psi)({\bf v}\cdot\nabla\psi)}{8m_1 m_2 \sinh{^2(\frac{1}{2}\beta\hbar\omega_1)}}\delta(\omega_1-\omega_2).
\label{69}
\end{equation}
The extra factor with frequency squared reflects the difference in dimension of the $\psi$ in interactions (\ref{50}) and (\ref{51}).

\section{Polarizable particles}
\label{sec6}

Although  interactions (\ref{15}) and (\ref{16}) give repulsive Casimir force, the friction force (\ref{69}) still has proper sign for a braking force. The magnitude of the force, however, has the same form as for the attractive situation. The only modification {besides a different $\psi$ is an extra $\omega_1^2$ factor that reflects} the $\zeta$ of expression (\ref{12}). Now result (\ref{69}) can be extended to a pair of polarizable magnetic and dielectric particles  following the development of Ref.~\cite{hoye12} whose equations will be designated with the numeral II. The main change is the form of the interaction that can be written in the form
\begin{equation}
-AF(t)=\psi_{ij}\dot{s}_{1i}s_{2j}, \quad \left(\sum\limits_{ij}\right).
\label{70}
\end{equation}
With (\ref{51}) there is also an $s_{1i}\dot{s}_{2j}$ term, but all weight can be put on only one of them as concluded below Eq.~(\ref{17}).

By introducing magnetic and electric polarizabilities $\alpha_1$ and $\alpha_2$ respectively, one has for  isotropic particles  $1/m_i=\omega_i^2\alpha_i$ as mentioned below Eq.~(II43). [This is consistent with the Fourier transform of the response function (II21) used there.  Note that the meaning of $\alpha$ is here and henceforth different from that in the previous sections.]
The friction force {(\ref{69})} in the $l$-direction then gets the form
\begin{equation}
F_{fl}=-G_{lq}v_q H\frac{\pi\beta\omega_1^2}{2}\delta(\omega_1-\omega_2),
\label{71}
\end{equation}
which is Eq.~(II43) with the additional factor $\omega_1^2$. The $H$ is given by (II40)
{
\begin{equation}
H=\frac{\hbar^2\omega_1\omega_2\alpha_1\alpha_2}{4\sinh{(\frac{1}{2}\beta \hbar\omega_1)}\sinh{(\frac{1}{2}\beta \hbar\omega_2)}}.
\label{72}
\end{equation}
}
The explicit form of the interaction follows from expression (\ref{15})
\begin{equation}
\psi_{ij}\dot{s}_{1i}s_{2j}=-\frac{1}{cr^2}(\dot{{\bf s}_1}\times\hat{r}){\bf s}_2=\frac{1}{cr^3}\varepsilon_{kij}x_k\dot{s}_{1i}s_{2j}
\label{73}
\end{equation}
(The symbol $\varepsilon_{kij}$ equals 1 for $kij$ in cyclic order, -1 in opposite order, and 0 otherwise.)
The gradient Eq.~(II26) of the interaction is needed
\begin{equation}
T_{lij}=\frac{\partial\psi_{ij}}{\partial x_l}=\frac{1}{c}\left(\frac{\delta_{lk}}{r^3}-\frac{3x_l x_k}{r^5}\right)\varepsilon_{kij}.
\label{74}
\end{equation}

Since the various components of the dipole moments are independent of each other ($\langle s_{li} s_{lj}\rangle=0$, $i\ne j$, $l=1,2$), as expressed by Eq.~(II19), one with Eq.~(\ref{74}) inserted {modifies Eq.~(II28) to
\begin{equation}
G_{lq}=T_{lij} T_{qij}= \frac{2}{c^2}\left(\frac{\delta_{lq}}{r^6}+\frac{3 x_l x_q}{r^8}\right).
\label{75}
\end{equation}
}
This inserted in expression (\ref{71}) gives the friction between the two oscillators. It extends Eq.~(II43) in a simple way to the present situation.

\section{Friction between a particle and a half-space and between two half-spaces}
\label{sec7}

The result for a pair of particles can be extended in a straightforward way to the situation with a particle and a half-space and to two parallel half-spaces by use of the equations of Ref.~\cite{hoye12}. Low particle density is assumed in this section by which forces are additive.

Assume that the half-space is located at $z \geq z_0$ such that its surface is parallel to the $xy$ plane at vertical position $z=z_0$. The dielectric particle, located at \tc{red}{$z=0$,} moves with constant velocity $v$ along the $x$ axis. Only $G_{11}$ is needed, and  with expression (\ref{75}) one finds
\begin{equation}
G_h=\rho\int\limits_{z>z_0} G_{11}\,dxdydz=\frac{\pi\rho}{2 c^2z_0^3}
\label{76}
\end{equation}
to replace result (II37). Here $\rho$ is the number density of particles.

For two parallel half-spaces with particle densities $\rho_1$ and $\rho_2$ and separated by a distance $d$ the friction force per unit area follows by use of Eq.~(II38) which in the present case gives
\begin{equation}
G=\rho_2\int\limits_d^\infty G_h\,dz_0=\frac{\pi}{4c^2d^2}\rho_1\rho_2.
\label{77}
\end{equation}

Altogether with Eq.~(\ref{71}) the friction force between a particle and a half-space becomes
\begin{equation}
F_h=-G_h v\omega_1^2 H\frac{\pi\beta}{2}\delta(\omega_1-\omega_2)
\label{78}
\end{equation}
and likewise between two half-spaces the friction force per unit area becomes
\begin{equation}
F=-G v\omega_1^2 H\frac{\pi\beta}{2}\delta(\omega_1-\omega_2)
\label{79}
\end{equation}
These results are the extension of results (II44) and (II45).  (The $v$ by mistake is missing in the reference.) Compared with the results of Ref.~\cite{hoye12},   factors $1/z_0^2$ or $1/d^2$ are replaced by the factor $(\omega_1/c)^2$ besides a slightly different numerical factor.

\section{General polarizability}
\label{sec8}

For a pair of oscillators with sharp frequencies $\omega_1$ and $\omega_2$ result (\ref{71}) along with results (\ref{75}), (\ref{78}), and (\ref{79}) are rather singular due to the $\delta$ function. However, for realistic oscillators with frequency spectra this singular behavior disappears. With Eqs.~(II46)-(II48) a polarizability $\alpha_{aK}=h(K^2)$ ($a=1,2$, $K=i\hbar\omega$) has a frequency spectrum $\alpha_a(m^2)$ such that \cite{hoye12,hoye82}
\begin{equation}
h(K^2)=\int\frac{\alpha_a(m^2)m^2}{K^2+m^2}\,d(m^2),
\label{80}
\end{equation}
\begin{equation}
\alpha_a(m^2) m^2=-\frac{1}{\pi}\Im[h(-m^2+i\gamma)],\quad m=\hbar\omega, \quad \gamma\rightarrow0+.
\label{81}
\end{equation}
Integrations over the frequencies give integrals (II49) and (II50) with an extra factor $\omega^2$ [$\int\delta(\omega_1-\omega_2)\,d(m_2^2)]d(m_1^2)=4(\hbar\omega_1)^2\,d\omega_1$]
\begin{equation}
H_0=\frac{\pi\beta}{2}\int\omega^2\frac{m^4\alpha_1(m^2)\alpha_2(m^2)}{\sinh^2(\frac{1}{2}\beta m)}\,d\omega.
\label{82}
\end{equation}
With this and Eqs.~(\ref{71}), (\ref{78}), and (\ref{79}) the various forces (II51) are replaced by
\begin{eqnarray}
\nonumber
F_{fl}&=&-G_{lq} v_q H_0,\\
F_h&=&-G_h vH_0,
\label{83}\\
\nonumber
F&=&-GvH_0.
\end{eqnarray}

\section{Friction at finite and zero temperature}
\label{sec9}

At $T=0$ the friction forces (\ref{83}), linear in $v$, vanish. However, for higher order in $v$ there will be non-zero friction also in this case as energy excitations with $\omega_2 \ne\omega_1$ become possible. To obtain it we follow the development of Ref.~\cite{hoye14} where energy dissipation is utilized with basis in the response function (\ref{52}). There this method was used both for finite and zero {temperatures,} and we do the same here to first recover the friction force (\ref{83}) with (\ref{82}) inserted for two half-spaces. Further to obtain explicit answers we use the frequency distribution for a metal for the dielectric half-space and assume it to be of the same form {also} for the magnetic one as a specific case.

The situation with two half-spaces is considered where Fourier transform is utilized in the $x$ and $y$ directions. Then $\nabla\rightarrow -i{\bf k}_\perp$, $d{\bf k}_\perp=dk_x dk_y$, ${\bf v}||{\bf k}_\perp$. With the present interaction (\ref{73}) one has ($k\rightarrow p$)
\begin{equation}
\psi_{ij}=-\frac{1}{c}\frac{\partial}{\partial x_p}\left(\frac{1}{r}\right)\varepsilon_{pij}
\label{84}
\end{equation}
with Fourier transform (in three dimensions)
\begin{equation}
c\tilde\psi_{ij}=ik_p\tilde\psi\varepsilon_{pij}, \quad \tilde\psi=\frac{4\pi}{k^2}.
\label{85}
\end{equation}
Likewise the transform of (\ref{74}) is
\begin{equation}
c\tilde T_{lij}=k_l k_p \tilde\psi\varepsilon_{pij}.
\label{86}
\end{equation}
The $G_{lq}$ of (\ref{75}) is integrated over ${\bf r}$-space. This can be converted to ${\bf k}$-space with use of both ${\bf k}$ and $-{\bf k}$ in the product of the two $\tilde T$. With (\ref{86}) this gives
\begin{equation}
c^2 \tilde G_{lq}=c^2\tilde T_{lij}\tilde T_{qij}=k_lk_qk_pk_m\tilde\psi^2\varepsilon_{pij}\varepsilon_{mij}=2k_lk_qk^2\tilde\psi^2.
\label{87}
\end{equation}
Now the transform should be limited to the $xy$-plane by which \cite{hoye14}
\begin{equation}
\tilde\psi\rightarrow\hat\psi(z_0,k_\perp)=\frac{2\pi e^{-q|z_0|}}{q}
\label{88}
\end{equation}
where $q=k_\perp$, $k_\perp^2=k_x^2+k_y^2$, $z_0=z_2-z_1$. This is Eq.~(III37) where the numeral III here and below designate
s the equations of Ref.~\cite{hoye14}. With $\pm ik_z=q$ for $z>0$ and $\pm ik_z=-q$ for $z<0$ one finds Eq.~(III40)
\begin{equation}
-ik_j ik_j=k_x^2+k_y^2+(\pm q)^2=k_\perp^2+q^2=2q^2.
\label{89}
\end{equation}
by which we {can write}
\begin{equation}
c^2\hat G_{11}=c^2\hat G_{xx}=\frac{1}{2}q^2\hat G(z_0,q),\quad  \hat G(z_0,q)=4q^2\hat\psi^2
\label{90}
\end{equation}
since by integration over orientations the average is $\langle k_x^2\rangle=(k_x^2+k_y^2)/2=q^2/2$. In the present case integrals (III43) and (III44) are modified to ($dk_xdk_y=2\pi q\,dq$)
{
\begin{equation}
\hat G(q)=\int\limits_{z_1>d,}\int\limits_{z_2<0}\hat G(z_0,q)\,dz_1dz_2=\frac{1}{c^2q^2}(2\pi)^2e^{-2qd},
\label{91}
\end{equation}
}
\begin{equation}
G=\frac{\rho_1\rho_2}{(2\pi)^2}\int\limits_0^\infty \frac{1}{2}q^2\hat G(q) 2\pi q\,dq=\frac{\pi\rho_1\rho_2}{4c^2 d^2},
\label{92}
\end{equation}
by which result (\ref{77}) is recovered.

To have an explicit result we may assume the Drude model for a metal to represent the dielectric half-space. Then with Eqs.~(III49)-(III51)
\begin{eqnarray}
\nonumber
\varepsilon=1+\frac{\omega_p^2}{\zeta(\zeta+\nu)}, \quad 2\pi\rho_1\alpha\rightarrow\frac{\varepsilon-1}{\varepsilon+1}\\
m^2\alpha_1(m^2)=D_1m, \quad D_1=\frac{\hbar\nu}{\rho_1(\pi\hbar\omega_p)^2}
\label{93}
\end{eqnarray}
for small $m$. The $\omega_p^2$ is the plasma frequency and $\nu$ is a constant related to the resistivity of the metal. The relation between polarizability $\alpha$ and $\varepsilon$, also valid for large $\varepsilon$, was established in Ref.~\cite{hoye13}. Now we here assume the magnetic half-space has a similar frequency spectrum
\begin{equation}
m^2\alpha_2(m^2)=D_2m
\label{94}
\end{equation}
With expression (\ref{82}) for $H_0$ instead of (III35) integral (III57) is modified into
{
\begin{equation}
H_0=\frac{\pi\beta}{2\hbar}D_1D_2\int\limits_0^\infty\frac{m^4\,dm}{\sinh^2(\frac{1}{2}\beta m)}=\frac{2\pi}{\beta^4\hbar}D_1D_2 I,
\label{95}
\end{equation}
}
\begin{equation}
I=\int\limits_0^\infty\frac{x^4 e^{-x}\,dx}{(1-e^{-x})^2}=\int_0^\infty \sum\limits_{n=1}^\infty x^4 n e^{-nx}\,dx=4!\sum\limits_{n=1}^\infty\frac{1}{n^4}=\frac{4\pi^4}{15}.
\label{96}
\end{equation}
With (\ref{83}) and (\ref{92}) this gives \tc{red}{the} friction force
\begin{equation}
F=-\frac{2\pi^6}{15}\left(\frac{d}{\beta c\hbar}\right)^2\frac{\rho_1\rho_2 D_1D_2}{\beta^2 d^4}\hbar v.
\label{97}
\end{equation}
Compared with result (III59) the main difference is the factor in parenthesis, and this factor is very small for reasonable values of $d$ and $\beta$.

To obtain the $T=0$ friction the response function (\ref{63}) {can} be written on form (III21) since $\omega_1\ne\omega_2$ is needed for $T=0$. With (\ref{64}) we find
\begin{eqnarray}
\nonumber
\phi(t)&=&C_-\sin(\omega_- t)+C_+\sin(\omega_+ t), \quad \omega_\pm=\vert\omega_1\pm\omega_2\vert,\\
C_\pm&=&\left(\frac{\omega_\mp}{2}\right)^2\frac{H}{\hbar}\sinh\left(\frac{1}{2}\beta\hbar\omega_\pm\right)
\label{98}
\end{eqnarray}
with $H$ given by (\ref{72}). The $C_-$ term determines the $T>0$ contribution just found above. For $T\rightarrow 0$ one finds (III47) modified into
\begin{equation}
C_+=\frac{1}{2}\left(\frac{\omega_-}{2}\right)^2\hbar\omega_1\omega_2\alpha_1\alpha_2.
\label{99}
\end{equation}
This modifies integral (III48) to
\begin{equation}
J(\omega_v)=2\pi\tau|\omega_v|\hbar^3\int\limits_0^{|\omega_v|}\left(\frac{\omega_-}{2}\right)^2\omega_1\omega_2 m_1 m_2\alpha_1(m_1^2)\alpha_2(m_2^2)\,d\omega_1
\label{100}
\end{equation}
with $\omega_1+\omega_2=\omega_+=|\omega_v|$, $\omega_v={\bf k}_\perp{\bf v}$. The $\tau$ is half the time for energy dissipation at velocity $v$. With frequency spectra (\ref{93}) and (\ref{94}) integral (III52) is replaced by
\begin{equation}
J(\omega_v)=2\pi\tau|\omega_v|\hbar^3D_1 D_2\frac{1}{4}\int\limits_0^{|\omega_v|}(\omega_1-\omega_2)^2\omega_1\omega_2\,d\omega_1=2\tau\omega_v^6H_P, \quad H_P=\frac{\pi}{120}\hbar^3D_1 D_2.
\label{101}
\end{equation}
By integration over orientations in the $xy$ plane {the integral above} (III53) is modified to ($\omega_v=k_x v$)
\begin{equation}
\int k_x^6\,d\phi=k_\perp^6\int\limits_0^{2\pi}\cos^6\phi\,d\phi=2\pi q^6\frac{5}{16}, \quad (q=k_\perp)
\label{102}
\end{equation}
such that $k_x^6$ can be replaced by $5q^6/16$ by which
\begin{equation}
J(\omega_v)=2\tau v^6 H_P\frac{5}{16}q^6.
\label{103}
\end{equation}

With Eq.~(\ref{91}) for $\hat G(q)$ integral (III54) is now modified to
\begin{equation}
G_P=\frac{\rho_1\rho_2}{(2\pi)^2}\int\limits_0^\infty\frac{5}{16}q^6\hat G(q)2\pi q\,dq=\frac{75\pi}{64c^2d^6}\rho_1\rho_2.
\label{104}
\end{equation}
In the present case the energy dissipation (III55) becomes $\Delta E_P=2\tau H_P v^6 G_P$ by which the friction force per unit area at temperature $T=0$ now replaces result (III56) with
\begin{equation}
F_P=-\frac{\Delta E}{2\tau v}=-\frac{5\pi^2}{512d^6}\left(\frac{v}{c}\right)^2\rho_1\rho_2 D_1 D_2(\hbar v)^3.
\label{105}
\end{equation}
Here the main difference from result (III56) is the very small factor  $(v/c)^2$. This is similar to the finite temperature result (\ref{97}) where another very small factor appeared for the same system with a magnetic half-space moving parallel to a dielectric one.
\begin{equation}
\label{}
\end{equation}

\section{Concluding remarks}

The basic microscopical system that we have analyzed, is an oscillating electric dipole transversely oriented with respect to an analogous magnetic dipole. This is extended to isotropic polarizable particles. In the quasistatic approximation the magnetic field created by an electric current according to Biot-Savart's law, is used. The quasistatic approximation implies that the static approximation used in most of our earlier works in this area \cite{hoye18,hoye92,hoye10,hoye12,hoye13,hoye14,hoye93,brevik88} can be generalized and extended to the magnetodielectric case in a natural way.
A characteristic property of the calculated force expressions is that they are heavily truncated relatively to the ordinary friction forces valid for purely dielectric media.

The very small friction relative to the case with both half-spaces dielectric can be understood from the magnetic field (\ref{13}). Compared with the electric field created by the same dipole moment it is reduced by a factor $r\zeta$. The characteristic length of the system is $r\sim d$, and at finite temperature the excitation energies to be transferred between the oscillators are $\hbar\omega\sim 1/\beta$ ($\zeta=i\omega/c$). Thus $r\zeta\sim d/(\beta c\hbar)$, the square of which appears in result (\ref{97}). Likewise for $T=0$ energy transfer can take place by excitations from the ground state with energies $\hbar\omega\sim \hbar\omega_v$, $\omega_v={\bf k}_\perp {\bf v}$, i.e.~$\omega\sim k_\perp v$. Further $rk_\perp\sim dk_\perp\sim 1$ by which $r\zeta\sim v/c$, the square of which appears in result (\ref{105}).

The corresponding results with both half-spaces dielectric have earlier been shown to be consistent with other approaches with full agreement or with some deviations restricted to numerical prefactors \cite{hoye13,hoye14,milton16}.

We have made use of quantum mechanical statistical methods throughout, similarly as in our earlier works. These methods turn out to be compact and effective.

\section*{Acknowledgment}
We acknowledge financial support from the Research Council of Norway, Project 250346.

\end{document}